\documentclass[longauth,letter]{aa}
\usepackage{graphicx}
\begin{document}
   \title{Hydrides in Young Stellar Objects: Radiation tracers in a protostar-disk-outflow system\thanks{{\it Herschel} is an ESA space observatory with science instruments provided by a European-led Principal Investigator consortia and with important participation from NASA}}
%\subtitle{Predictions and Observations}
\author{A.O. Benz\inst{\ref{inst1}}
\and S. Bruderer\inst{\ref{inst1}}
\and E.F. van Dishoeck\inst{\ref{inst2},\ref{inst3}}
\and P. St\"auber\inst{\ref{inst1}}
\and S.F.~Wampfler\inst{\ref{inst1}}
\and M. Melchior\inst{\ref{inst1},\ref{inst34}}
\and C.~Dedes\inst{\ref{inst1}}
\and F.~Wyrowski\inst{\ref{inst30}}
\and S.D.~Doty\inst{\ref{inst19}}
\and F.~van der Tak\inst{\ref{inst10},\ref{inst11}}
\and W. B\"achtold\inst{\ref{inst35}}
\and A. Csillaghy\inst{\ref{inst34}}
\and A. Megej\inst{\ref{inst35}}
\and C. Monstein\inst{\ref{inst1}}
\and M. Soldati\inst{\ref{inst34}}
\and R.~Bachiller\inst{\ref{inst12}}
\and A.~Baudry\inst{\ref{inst6}}
\and M.~Benedettini\inst{\ref{inst13}}
\and E.~Bergin\inst{\ref{inst14}}
\and P.~Bjerkeli\inst{\ref{inst9}}
\and G.A.~Blake\inst{\ref{inst15}}
\and S.~Bontemps\inst{\ref{inst6}}
\and J.~Braine\inst{\ref{inst6}}
\and P.~Caselli\inst{\ref{inst4},\ref{inst5}}
\and J.~Cernicharo\inst{\ref{inst16}}
\and C.~Codella\inst{\ref{inst5}}
\and F.~Daniel\inst{\ref{inst16}}
\and A.M.~di~Giorgio\inst{\ref{inst13}}
\and P. Dieleman\inst{\ref{inst10}}
\and C.~Dominik\inst{\ref{inst17},\ref{inst18}}
\and P.~Encrenaz\inst{\ref{inst20}}
\and M.~Fich\inst{\ref{inst21}}
\and A.~Fuente\inst{\ref{inst22}}
\and T.~Giannini\inst{\ref{inst23}}
\and J.R.~Goicoechea\inst{\ref{inst16}}
\and Th.~de~Graauw\inst{\ref{inst10}}
\and F.~Helmich\inst{\ref{inst10}}
\and G.J.~Herczeg\inst{\ref{inst3}}
\and F.~Herpin\inst{\ref{inst6}}
\and M.R.~Hogerheijde\inst{\ref{inst2}}
\and T.~Jacq\inst{\ref{inst5}}
\and W. Jellema\inst{\ref{inst10}}
\and D.~Johnstone\inst{\ref{inst7},\ref{inst8}}
\and J.K.~J{\o}rgensen\inst{\ref{inst24}}
\and L.E.~Kristensen\inst{\ref{inst2}}
\and B.~Larsson\inst{\ref{inst25}}
\and D.~Lis\inst{\ref{inst26}}
\and R.~Liseau\inst{\ref{inst9}}
\and M.~Marseille\inst{\ref{inst10}}
\and C.~M$^{\textrm c}$Coey\inst{\ref{inst21},\ref{inst27}}
\and G.~Melnick\inst{\ref{inst28}}
\and D.~Neufeld\inst{\ref{inst29}}
\and B.~Nisini\inst{\ref{inst23}}
\and M.~Olberg\inst{\ref{inst9}}
\and V. Ossenkopf\inst{\ref{inst43}}
\and B.~Parise\inst{\ref{inst30}}
\and J.C.~Pearson\inst{\ref{inst31}}
\and R.~Plume\inst{\ref{inst32}}
\and C.~Risacher\inst{\ref{inst10}}
\and J.~Santiago-Garc\'{i}a\inst{\ref{inst33}}
\and P.~Saraceno\inst{\ref{inst13}}
\and R. Schieder\inst{\ref{inst43}}
\and R.~Shipman\inst{\ref{inst10}}
\and J. Stutzki\inst{\ref{inst43}}
\and M.~Tafalla\inst{\ref{inst12}}
\and A.G.G.M. Tielens\inst{\ref{inst2}}
\and T.A.~van~Kempen\inst{\ref{inst28}}
\and R.~Visser\inst{\ref{inst2}}
\and U.A.~Y{\i}ld{\i}z\inst{\ref{inst2}}
}

\institute{
Institute of Astronomy, ETH Zurich, 8093 Zurich, Switzerland\label{inst1}
\and
Leiden Observatory, Leiden University, PO Box 9513, 2300 RA Leiden, The Netherlands\label{inst2}
\and
Max Planck Institut f\"{u}r extraterrestrische Physik, Giessenbachstrasse 1, 85748 Garching, Germany\label{inst3}
\and
School of Physics and Astronomy, University of Leeds, Leeds LS2 9JT, UK\label{inst4}
\and
INAF - Osservatorio Astrofisico di Arcetri, Largo E. Fermi 5, 50125 Firenze, Italy\label{inst5}
\and
Universit\'{e} de Bordeaux, Laboratoire d'Astrophysique de Bordeaux, France; CNRS/INSU, UMR 5804, Floirac, France\label{inst6}
\and
National Research Council Canada, Herzberg Institute of Astrophysics, 5071 West Saanich Road, Victoria, BC V9E 2E7, Canada\label{inst7}
\and
Department of Physics and Astronomy, University of Victoria, Victoria, BC V8P 1A1, Canada\label{inst8}
\and
Department of Radio and Space Science, Chalmers University of Technology, Onsala Space Observatory, 439 92 Onsala, Sweden\label{inst9}
\and
SRON Netherlands Institute for Space Research, PO Box 800, 9700 AV, Groningen, The Netherlands\label{inst10}
\and
Kapteyn Astronomical Institute, University of Groningen, PO Box 800, 9700 AV, Groningen, The Netherlands\label{inst11}
\and
Observatorio Astron\'{o}mico Nacional (IGN), Calle Alfonso XII,3. 28014, Madrid, Spain\label{inst12}
\and
INAF - Istituto di Fisica dello Spazio Interplanetario, Area di Ricerca di Tor Vergata, via Fosso del Cavaliere 100, 00133 Roma, Italy\label{inst13}
\and
Department of Astronomy, The University of Michigan, 500 Church Street, Ann Arbor, MI 48109-1042, USA\label{inst14}
\and
California Institute of Technology, Division of Geological and Planetary Sciences, MS 150-21, Pasadena, CA 91125, USA\label{inst15}
\and
Centro de Astrobiolog\'{\i}a. Departamento de Astrof\'{\i}sica. CSIC-INTA. Carretera de Ajalvir, Km 4, Torrej\'{o}n de Ardoz. 28850, Madrid, Spain.\label{inst16}
\and
Astronomical Institute Anton Pannekoek, University of Amsterdam, Kruislaan 403, 1098 SJ Amsterdam, The Netherlands\label{inst17}
\and
Department of Astrophysics/IMAPP, Radboud University Nijmegen, P.O. Box 9010, 6500 GL Nijmegen, The Netherlands\label{inst18}
\and
Department of Physics and Astronomy, Denison University, Granville, OH, 43023, USA\label{inst19}
\and
LERMA and UMR 8112 du CNRS, Observatoire de Paris, 61 Av. de l'Observatoire, 75014 Paris, France\label{inst20}
\and
University of Waterloo, Department of Physics and Astronomy, Waterloo, Ontario, Canada\label{inst21}
\and
Observatorio Astron\'{o}mico Nacional, Apartado 112, 28803 Alcal\'{a} de Henares, Spain\label{inst22}
\and
INAF - Osservatorio Astronomico di Roma, 00040 Monte Porzio catone, Italy\label{inst23}
\and
Centre for Star and Planet Formation, Natural History Museum of Denmark, University of Copenhagen,
{\O}ster Voldgade 5-7, DK-1350 Copenhagen K., Denmark\label{inst24}
\and
Department of Astronomy, Stockholm University, AlbaNova, 106 91 Stockholm, Sweden\label{inst25}
\and
California Institute of Technology, Cahill Center for Astronomy and Astrophysics, MS 301-17, Pasadena, CA 91125, USA\label{inst26}
\and
the University of Western Ontario, Department of Physics and Astronomy, London, Ontario, N6A 3K7, Canada\label{inst27}
\and
Harvard-Smithsonian Center for Astrophysics, 60 Garden Street, MS 42, Cambridge, MA 02138, USA\label{inst28}
\and
Department of Physics and Astronomy, Johns Hopkins University, 3400 North Charles Street, Baltimore, MD 21218, USA\label{inst29}
\and
Max-Planck-Institut f\"{u}r Radioastronomie, Auf dem H\"{u}gel 69, 53121 Bonn, Germany\label{inst30}
\and
Jet Propulsion Laboratory, California Institute of Technology, Pasadena, CA 91109, USA\label{inst31}
\and
Department of Physics and Astronomy, University of Calgary, Calgary, T2N 1N4, AB, Canada\label{inst32}
\and
Instituto de Radioastronom\'{i}a Milim\'{e}trica (IRAM), Avenida Divina Pastora 7, N\'{u}cleo Central, E-18012 Granada, Spain\label{inst33}
\and
%Department of Earth and Planetary Sciences, Kobe University, Nada, Kobe 657-8501, Japan\label{inst34}
%\and
%Universit\'{e} Pierre et Marie Curie, LPMAA UMR CNRS 7092, Case 76, 4 place Jussieu, 75252 Paris Cedex 05, France\label{inst35}
%\and
%Observatoire de Paris-Meudon, LUTH UMR CNRS 8102, 5 place Jules Janssen, 92195 Meudon Cedex, France\label{inst36}
%\and
%Department of Physics and Astronomy, San Jose State University, One Washington Square, San Jose, CA 95192, USA\label{inst37}
%\and
%Laboratoire d'Astrophysique de Grenoble, CNRS/Universit\'{e} Joseph Fourier (UMR5571) BP 53, F-38041 Grenoble cedex 9, %France\label{inst38}
%\and
%European Southern Observatory, Karl-Schwarzschild-Str. 2, 85748 Garching, Germany\label{inst39}
%\and
%Department of Physics, The University of Tokyo, Hongo, Bunkyo-ku, Tokyo 113-0033, Japan\label{inst40}
%\and
%Department of Physics and Astronomy, University College London, Gower Street, London WC1E6BT\label{inst41}
%\and
%Department of Physics, The University of Tokyo, Hongo, Bunkyo-ku, Tokyo 113-0033, Japan\label{inst42}
%\and
KOSMA, I. Physik. Institut, Universit\"{a}t zu K\"{o}ln, Z\"{u}lpicher Str. 77, D 50937 K\"{o}ln, Germany\label{inst43}
%\and
%California Institute of Technology, 1200 E. California Bl., MC 100-22, Pasadena, CA. 91125  USA\label{inst44}
%\and
%Experimental Physics Dept., National University of Ireland Maynooth, Co. Kildare. Ireland\label{inst45}
\and
Institute of 4D Technologies, University of Applied Sciences NW, CH-5210 Windisch, Switzerland\label{inst34}
\and
Laboratory for Electromagnetic Fields and Microwave Electronics, ETH Zurich, 8092 Zurich, Switzerland\label{inst35}
}

\authorrunning{A.O. Benz et al.}
\titlerunning{Radiation tracers in YSOs}
   \date{Received May 31, 2010; accepted xxxx}

\abstract
% context heading (optional)
%{} leave it empty if necessary - but it must be there
{Hydrides of the most abundant heavier elements are fundamental molecules in cosmic chemistry. Some of them trace gas irradiated by UV or X-rays.}
% aims heading (mandatory)
{We explore the abundances of major hydrides in W3 IRS5, a prototypical region of high-mass star formation.}
% methods heading (mandatory)
{W3 IRS5 was observed by HIFI on the {{\it Herschel}} Space Observatory with deep integration ($\simeq$ 2500 s) in 8 spectral regions. }
% results heading (mandatory)
{The target lines including CH, NH, H$_3$O$^+$, and the new molecules SH$^+$, H$_2$O$^+$, and OH$^+$ are detected. The H$_2$O$^+$ and OH$^+$ $J=1-0$ lines are found mostly in absorption, but also appear to exhibit weak emission (P-Cyg-like). Emission requires high density, thus originates most likely near the protostar. This is corroborated by the absence of line shifts relative to the young stellar object (YSO). In addition, H$_2$O$^+$ and OH$^+$ also contain strong absorption components at a velocity shifted relative to W3 IRS5, which are attributed to foreground clouds.}
% conclusions
{The molecular column densities derived from observations correlate well with the predictions of a model that assumes the main emission region is in outflow walls, heated and  irradiated by protostellar UV radiation.}


   \maketitle
%
%________________________________________________________________

\section{Introduction}

In interstellar clouds, chemical reactions with hydrogen molecules lead to an elementary class of molecules that represent key species in the chemical evolution to larger molecules. These fundamental molecules, known as hydrides, include OH, CH, NH, SH, H$_2$O, and their ions, OH$^+$, CH$^+$, NH$^+$, SH$^+$, H$_2$O$^+$, and H$_3$O$^+$. The combination of hydrogen atoms with a heavier atom causes large dipole moments and large rotation constants, particularly in diatomic hydrides. This widely separates the excitation levels. Only low-J lines are excited at temperatures relevant to star and planet formation. These lines have now become observable with the {\it Herschel} Space Observatory (Pilbratt et al. 2010).

Many hydrides have a high activation energy in their formation paths. However, if high-energy photons - far UV (FUV) or X-rays - interact with the molecular gas and heat it, hydrides and particularly their ions are greatly enhanced in abundance (e.g., Hollenbach \& Tielens 1999). Ionized hydrides are chemically even more active and can substantially drive chemical evolution. Many of the above hydrides have been observed from the ground in absorption in diffuse interstellar clouds (e.g. Swings \& Rosenfeld 1937; Menten et al. 2010; Wyrowski et al. 2010a).

Here we focus on hydrides in dense star-forming regions surrounding protostars. We report first results of the `Radiation Diagnostics' project within the Herschel key program `Water In Star-forming regions with {\it Herschel}' (WISH, van Dishoeck et al. 2010). The `Radiation Diagnostics' subprogram aims to trace high-energy radiation through the most abundant hydrides and their ions and relate them to the chemical network of water. Preparing for this project, St\"auber et al. (2005, 2006) showed that hydrides may be enhanced in star-forming regions affected by far UV and/or X-ray irradiation. In a second step, optimal lines were selected using catalogued data in the available databases or computed from measured molecular constants (Bruderer 2006).

\begin{table*}[htb]
\begin{center}
\resizebox{15cm}{!}{
\begin{tabular}{lcrcc|rrrrrr}   % c = center, l = left justified, r=right just.
\hline \hline
Mole- &Tran-& Frequency&$E_u$&$A_{ul}$&Line peak&Line & Line &Line flux&Absorption&Column \\
cule &sition& & & &T$_{mb}$&width&shift &$\int T_{mb} dv$&$\int \tau dv$&density \\
& & [GHz]&[K]&[s$^{-1}$]&[mK]&[km s$^{-1}$]&[km s$^{-1}$]&[K km s$^{-1}$]&[km s$^{-1}$]&[cm$^{-2}$]\\
\hline\\
CH&$1_{-1}-1_{1}$ &536.7611$^a$& 25.76& 6.4(-4)& 740$\pm$7&10.6 &+ 1.6 &14.8$\pm$0.02& &1.3(13) \\
NH&$1_1$-$0_1$  &999.9734$^a$& 47.99& 5.2(-2)&- 446$\pm$13 &2.7 &- 0.8 &&3.20$\pm$0.02 &4.0(13)\\
SH&3$_{1}$ - 2$_{-1}$&1447.0123$^a$&640.6& 8.1(-3)& $<$220& & & $<$ 0.98&&$<$ 5.0(11) \\
OH$^+$&$1_1-0_1$&1033.1186$^a$ & 49.58& 1.8(-2)&- 790$\pm$21 &5.9  &+34.8&& 14.6$\pm$0.03&7.1(13)  \\
OH$^+$&$2_1-1_1$ &1892.2271$^a$ & 140.4& 5.9(-2) &$<$255 & & &$<$1.13&  &$<$ 1.1(11) \\
%NH$^+$&$N_{JP}=1_{\frac{3}{2}+}-1_{\frac{1}{2}-}$ &1012.5400$^d$& 48.59& 5.4(-2)& $<$& & &$<$&  \\
NH$^+$&$1_{\frac{3}{2}-}-1_{\frac{1}{2}+}$&1019.2107$^d$ & 48.91& 5.5(-2)& $<$0.032& & &$<$ 0.14& &$<$5.2(9) \\
SH$^+$&$1_{2\frac{5}{2}}-0_{1{\frac{3}{2}}}$&526.0479$^a$& 25.25& 9.7(-4)&65$\pm$3&4.4 &- 0.3 & 0.73$\pm$0.1&&4.1(11) \\
SH$^+$&2$_{3\frac{7}{2}}$ - 1$_{2\frac{5}{2}}$&1082.9117$^a$& 77.2&9.1(-2)&$<$37 &  & &$<$ 0.17 & &$<$ 3.9(10)\\
SH$^+$&3$_{4\frac{9}{2}}$ - 2$_{3\frac{7}{2}}$&1632.5179$^a$& 155.6&3.1(-2)&$<$151 &  & &$<$ 0.68& &$<$ 9.8(10) \\
H$_2$O$^+$&$3_{12}-3_{03}$&999.8213$^c$& 223.9& 2.3(-2)$^c$& $<$46& & & $<$ 0.20& &$<$ 1.7(10)\\
H$_2$O$^+$&$1_{11\frac{3}{2}}-0_{00\frac{1}{2}}$ &1115.2040$^b$ & 53.52& 3.1(-2)$^c$&- 285$\pm$24 &5.1&+39.1& &2.21$\pm$0.03& 4.6(12) \\
H$_3$O$^+$&4$_{30}$ - 3$_{31}$&1031.2937$^e$&232.2 & 5.1(-3)&570$\pm$10& 6.2&+0.5&3.8$\pm$0.3& &9.7(11) \\
H$_3$O$^+$&4$_{20}$ - 3$_{21}$ &1069.8266$^e$&268.8&9.8(-3)&230$\pm$10 &4.7 &- 0.3 &1.30$\pm$0.03& &2.9(11) \\
H$_3$O$^+$&6$_{21}$ - 6$_{20}$ &1454.5625$^e$&692.6&7.1(-3)&$<$ 245&&  & $<$ 0.66& &$<$ 3.8(11) \\
H$_3$O$^+$&2$_{11}$ - 2$_{10}$ &1632.0910$^e$&143.1& 1.7(-2)&145$\pm$46  &6.7&+0.3  &1.20$\pm$0.12& &3.7(11) \\
\hline
\end{tabular}}
\end{center}
\begin{flushleft}
\footnotesize{Molecular data are taken from: $^a$ CDMS (M\"uller et al. 2001), $^b$M\"urtz et al. (1998), $^c$Bruderer (2006), $^d$H\"ubers et al. (2009)}, $^e$JPL catalogue
\end{flushleft}
\vskip-0.1cm
\caption{Frequency, upper level energy, and Einstein coefficient of molecules and lines observed by {\it Herschel/HIFI} towards W3 IRS5. The numbers in parentheses give the decimal power. Negative peak fluxes signify background minus line temperature of  lines in absorption. Line widths refer to the FWHP value of the most intense line peak or absorption and its line shift to the systemic velocity of -38.4 km s$^{-1}$. Non-detected peak fluxes are 5$\sigma$ upper limits at 1 km s$^{-1}$ resolution, and non-detected line fluxes assume a 5 km s$^{-1}$ line width. The column densities refer to the upper (emission) or lower (absorption) state.}
\label{table}
\end{table*}

Spherically symmetric models including FUV and X-rays have found that most hydrides would be too weak for {\it Herschel} detection. However, evidence of extended high-energy irradiation (Doty et al. 2004) and in particular the observation of CO$^+$, with an abundance four orders of magnitude larger than predicted, in W3 IRS5 and other YSOs with the JCMT (St\"auber et al. 2007) raised the expectation that the effects of irradiation may be more dramatic in asymmetric reality. Bruderer et al. (2009) modeled and interpreted the CO$^+$ anomaly as FUV radiation originating in the YSOs and irradiating the walls of outflow cavities.

Based on these predictions, the `Radiation Diagnostics' observations started with deep integrations of a large number of hydrides that was to be followed by a survey of a few species in many sources of various ages and masses. Here we present exploratory observations towards W3 IRS5, a nearby region (1.83 kpc, Imai et al. 2000) of high-mass star formation, moving at -- 38.4 km s$^{-1}$ in the local standard of rest (LSR). W3 IRS5 resembles the Trapezium cluster in Orion but is considerably younger. At least six radio sources are in the {\it Herschel} beam. They represent `hypercompact' H II regions produced by high-mass YSOs, at least two of which are O stars (van der Tak, Tuthill, \& Danchi 2005; Rod\'on et al. 2008). In this paper, we report the observations of major hydrides towards W3 IRS5 apart from H$_2$O, CH$^+$, and OH, which are studied in other projects of WISH (Chavarr\'{\i}a et al. 2010; Wampfler et al. in prep.).

\section{Observations}
The {\it Heterodyne Instrument for the Far Infrared} (HIFI, de Graauw et al. 2010) on {\it Herschel} observed W3 IRS5 between 1 and 8 March 2010 in the Science Demonstration Phase in eight 4 GHz frequency bands for about 2500 s each. One of them includes [C II] at 1900.5369 GHz. We used the Wide Band Spectrometer which has a spectral resolution of 1.1 MHz, and HIPE 3.0 for pipeline and data analysis. The data were taken by double beam switching (DBS), the high-frequency (HEB) bands in fast DBS mode. The off-source position was at a distance of 3 arcmin in the NE and SW without remarkable IR sources.

The current accuracy of the velocity calibration is estimated to be better than 2 km s$^{-1}$. The antenna temperature was converted to main beam temperature, using pre-flight antenna efficiencies. After visual inspection and defringing, the V polarization was shifted linearly in flux to match the H polarization. The two polarizations were then added. A second observation of equal length was made using a local oscillator frequency shifted by 10 km s$^{-1}$. The two data sets were plotted in both upper and lower sideband presentation. If a line matched velocities in one sideband and was double with 20 km s$^{-1}$ separation in the other, the frequency of the former was assumed. All lines of interest could be attributed to a sideband without ambiguity. The continuum was divided by 2 for double sideband observations, assuming that it is the same in both sidebands, of equal sensitivity.

\begin{figure*}[t]
\centering
\begin{minipage}[t]{0.3\linewidth}
\centering
\includegraphics[width=\linewidth]{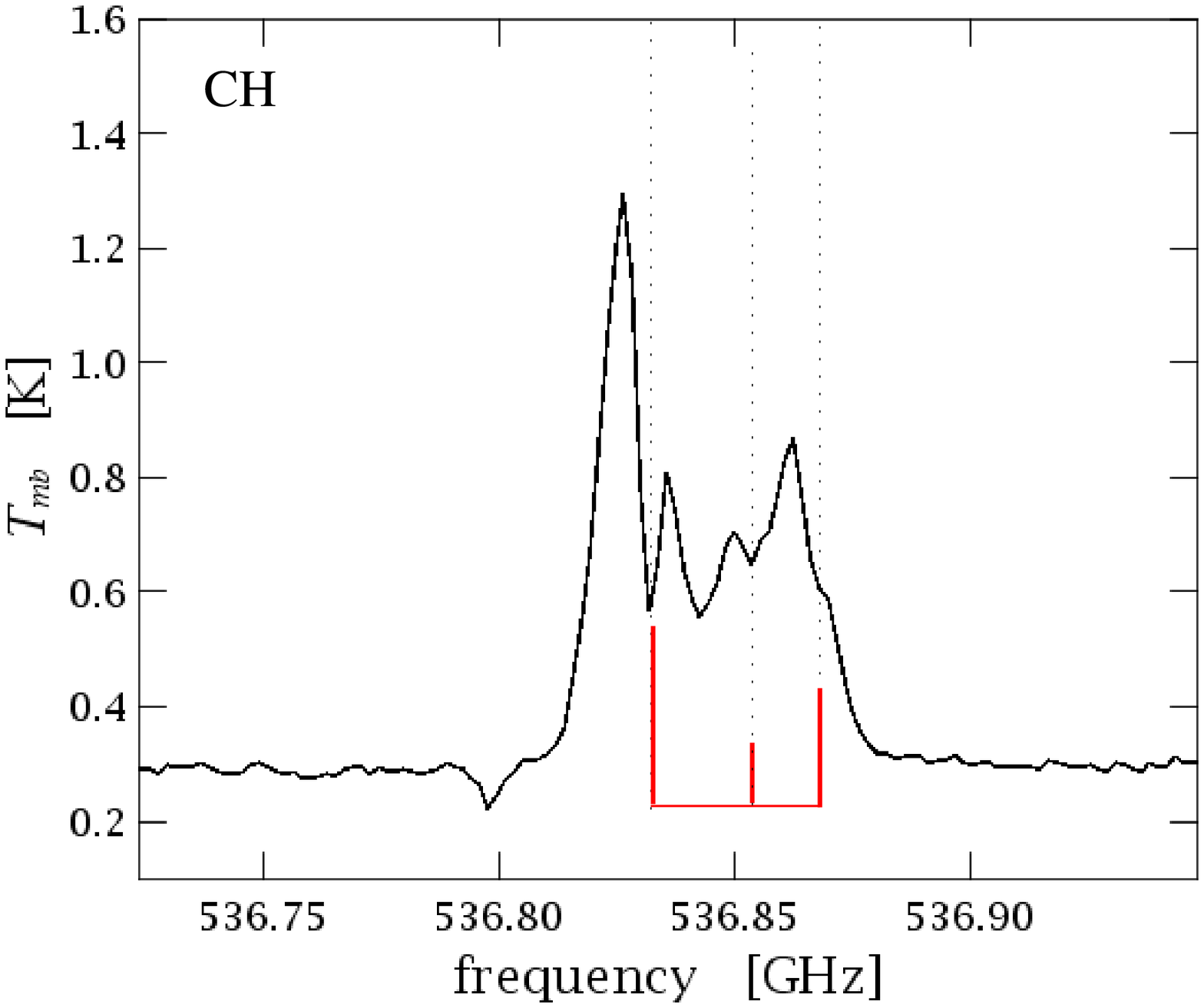}
\end{minipage}
\hspace{-0.2cm}
\centering
\begin{minipage}[t]{0.3\linewidth}
\centering
\includegraphics[width=\linewidth]{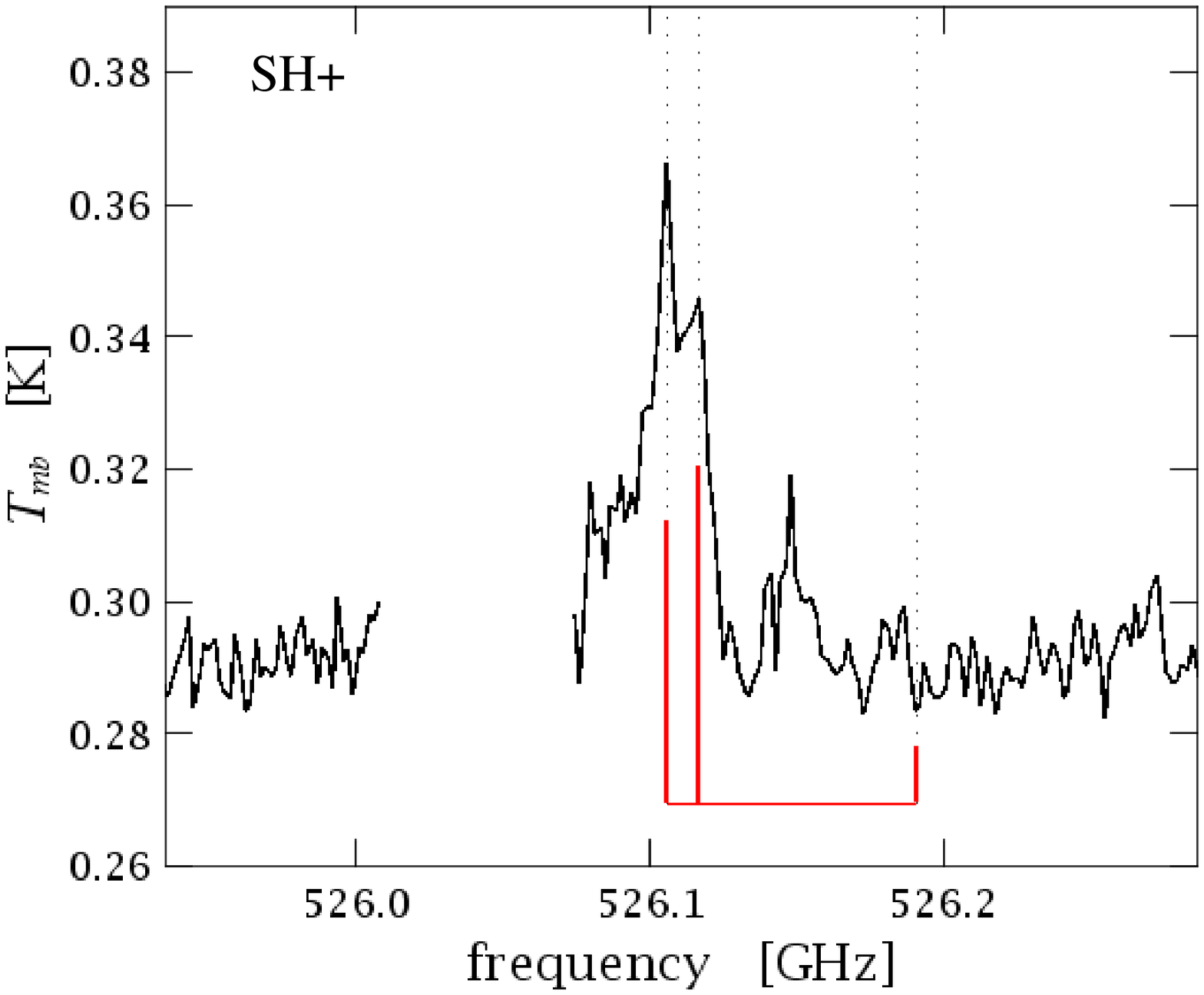}
\end{minipage}
\hspace{-0.2cm}
\begin{minipage}[t]{0.3\linewidth}
\centering
\includegraphics[width=\linewidth]{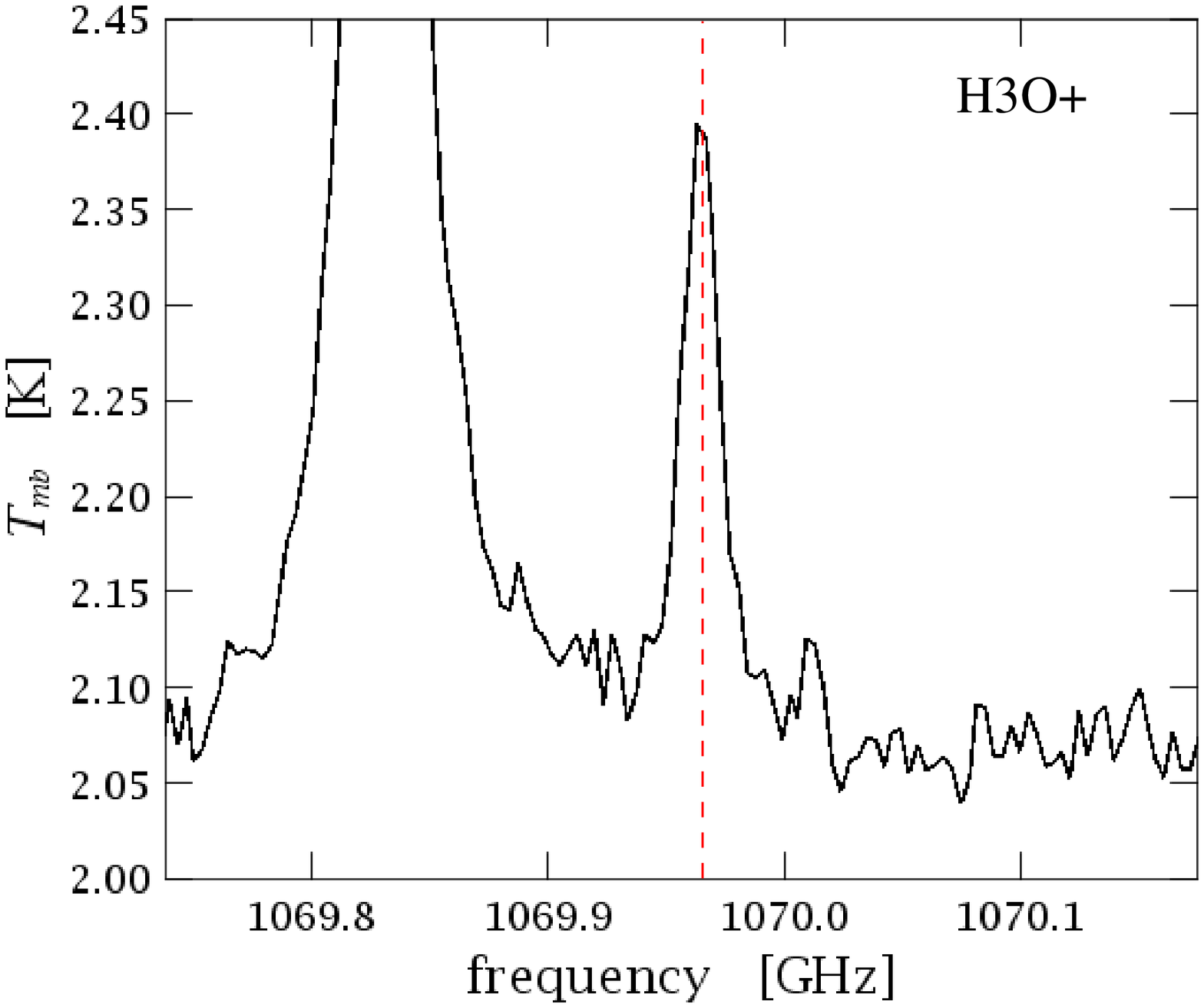}
\end{minipage}\\[-5pt]
\centering
\begin{minipage}[t]{0.3\linewidth}
\centering
\includegraphics[width=\linewidth]{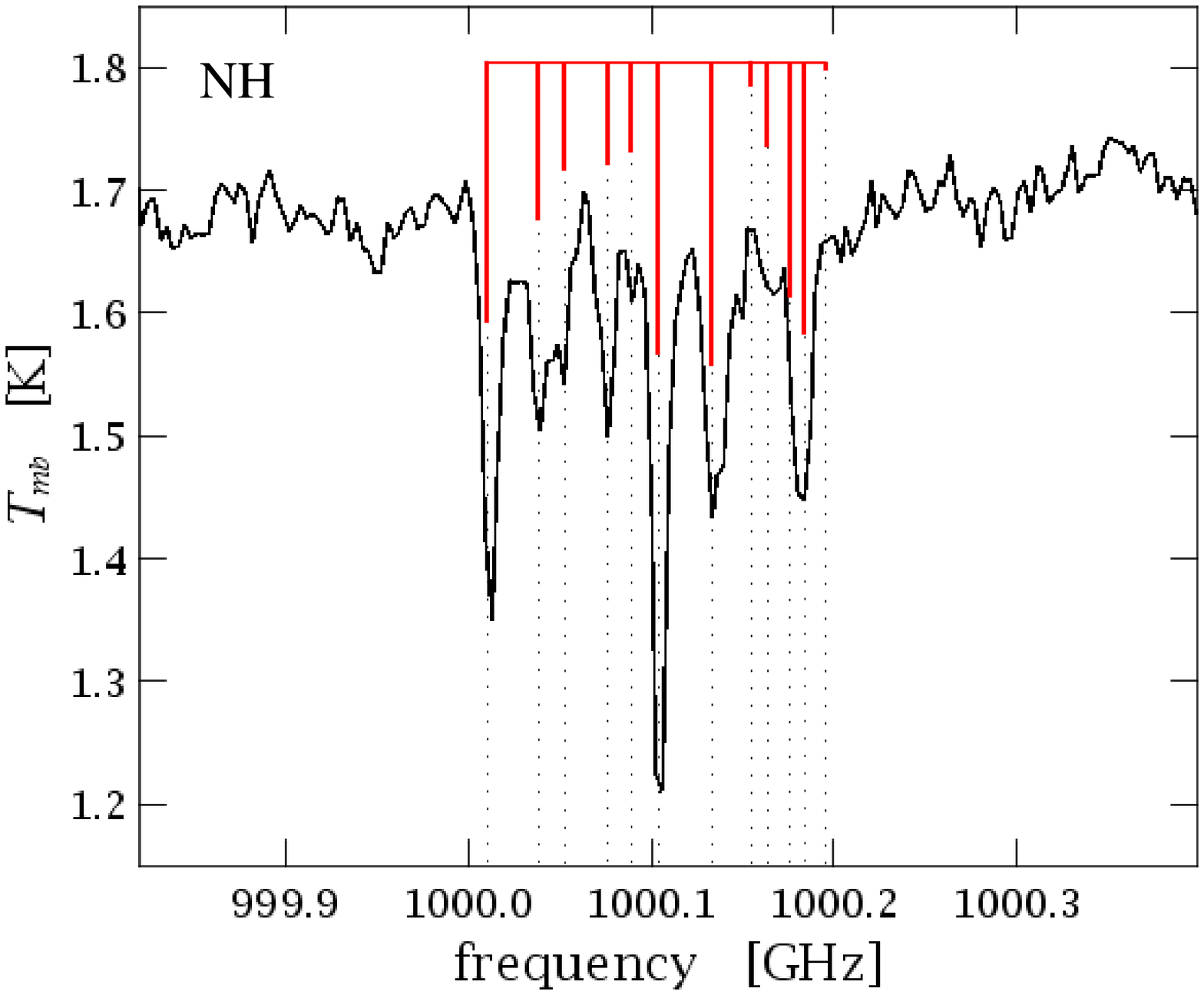}
\end{minipage}
\hspace{-0.2cm}
\begin{minipage}[t]{0.3\linewidth}
\centering
\includegraphics[width=\linewidth]{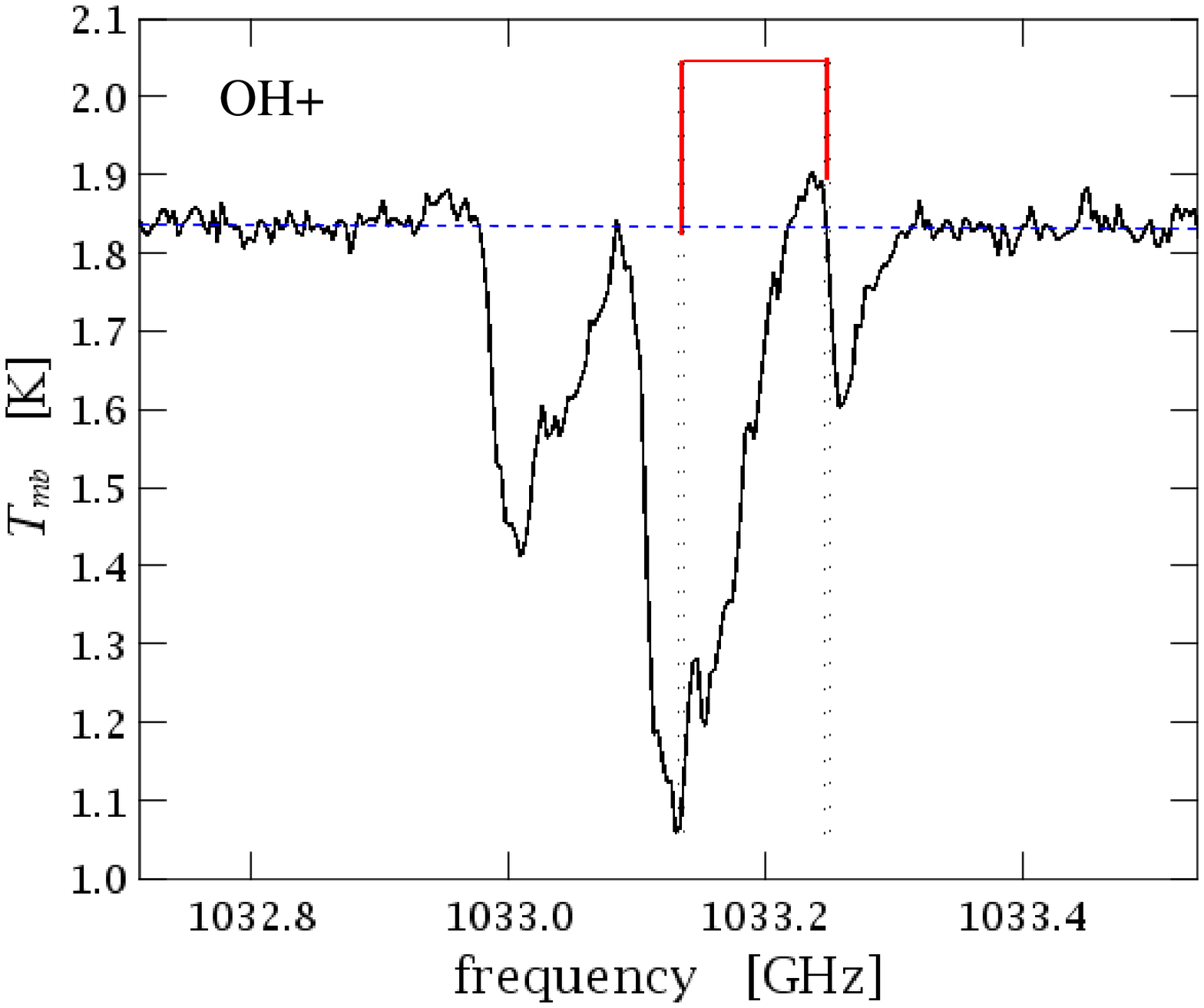}
\end{minipage}
\hspace{-0.2cm}
\begin{minipage}[t]{0.3\linewidth}
\centering
\includegraphics[width=\linewidth]{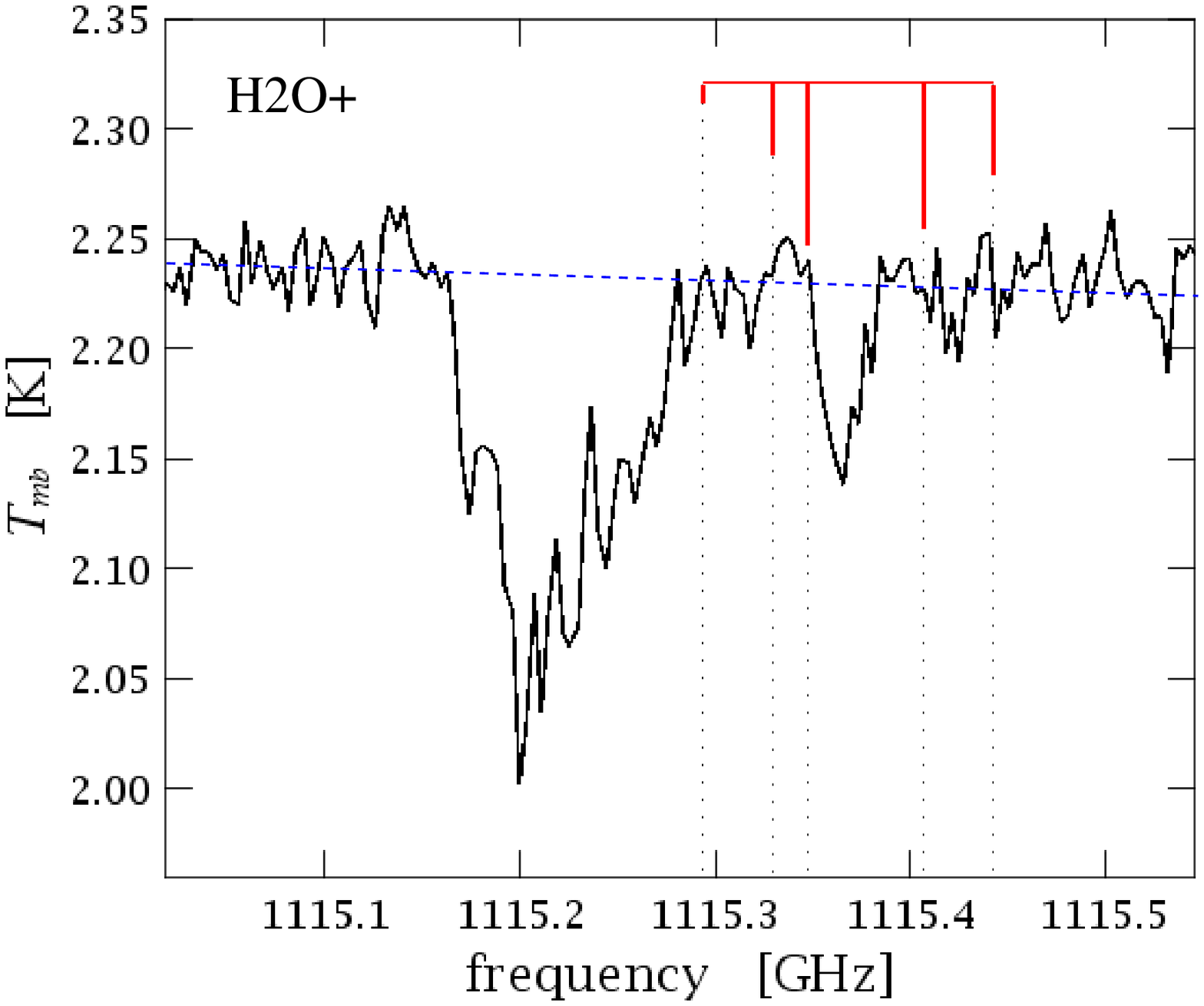}
\end{minipage}
\caption{Spectral lines observed towards W3 IRS5 in main beam temperature. The positions of the theoretical fine{\bf/hyperfine}  structure lines are shown shifted by -38.4 km s$^{-1}$, the systemic motion of the YSO. The length of the {\bf vertical} bars indicates the theoretical intensities in arbitrary scale. The frequency binning is 1 km s$^{-1}$.}
\end{figure*}

\section{Results}

Most lines (except H$_3$O$^+$) are split by fine or hyperfine interaction as indicated in Fig. 1. Table 1 lists the observed lines (strongest only for multiples) and summarizes the quantitative observational results. Molecules here detected for the first time in star-forming regions include H$_2$O$^+$, OH$^+$, and SH$^+$. Having the most prominent line near the H$_2$O para ground-state line at 1113.3 GHz, H$_2$O$^+$ is serendipitously detected in many {\it Herschel} observations. H$_2$O$^+$ and OH$^+$ are detected in absorption by the interstellar medium (Bruderer et al. 2010b; Ceccarelli et al. 2010; Gerin et al. 2010; Falgarone et al. 2010; Neufeld et al. (2010); Ossenkopf et al. 2010; Schilke et al. 2010), but also near the systemic velocity of other high-mass YSOs (Wyrowski et al. 2010b).

Several lines are found in absorption, indicated in Table 1 by negative peak values relative to the continuum. All lines predominantly in absorption (NH, OH$^+$, H$_2$O$^+$) originate in molecules in the ground state with a $J=1$ level energy exceeding 47 K. Lines of CH and SH$^+$, observed in emission, are transitions from a $J=1$ level of energy less than 26 K. We do not detect SH and NH$^+$ above the 5$\sigma$ limit.

The observed line shifts are within $\pm$2 km s$^{-1}$ of the YSO, thus within the accuracy of the frequency calibration, spectral resolution, and molecular data. The only deviations in Table 1 are OH$^+$ and H$_2$O$^+$. Their strongest peaks, both in absorption, are shifted by 34 - 39 km s$^{-1}$ relative to the YSO, similar to an absorption feature in [C II].

Figure 2 illustrates a possible explanation of the exceptional line shifts of OH$^+$ and H$_2$O$^+$. In both cases, a theoretically strong hyperfine structure line corresponds to a smaller absorption dip within less than 5 km s$^{-1}$ of the YSO. For OH$^+$, the low-frequency line (a doublet) is blended with the high-frequency line of another, red-shifted OH$^+$ component. We thus interpret the spectra of OH$^+$ and H$_2$O$^+$ as the superimposition of a component originating in the star-forming region and another component at lower frequency related to an outflow or the foreground interstellar medium near zero velocity in the LSR.

The line widths in Table 1 are between 4 and 7 km s$^{-1}$, measured at the component with the systemic velocity of the YSO. Two lines attract attention: ({\it i}) CH has an absorption feature near the YSO velocity. This absorption reduces the peak flux and widens the line profile at half power. ({\it ii}) The line width of NH is extremely narrow in all hyperfine structure components.

The lines are generally symmetric. A remarkable exception is OH$^+$, showing blue tails. We note that the shifted lines of OH$^+$ (unblended line) and H$_2$O$^+$ are 17.7 and 11.1 km s$^{-1}$ wide, respectively, indicating that they have a different origin from the component at zero systemic velocity.

Several lines in Fig. 1 show slightly blue-shifted absorption and red-shifted emission, thus a P-Cyg-like behavior. Figure 2 compares them with the [C II] line. CH (strongest component) exhibits similar absorption to [C II], their peak absorptions being at -1.6 km s$^{-1}$ and -1.7 km s$^{-1}$, respectively, relative to the YSO motion. OH$^+$ and H$_2$O$^+$ differ, having absorption peaks at -2.5 and -4.8 km s$^{-1}$, respectively, and  emission peaks at 4.2 and 3.5 km s$^{-1}$.

Column densities are given in Table 1 for the upper energy level of transitions in emission, and for the lower energy level of lines in absorption. The fine/hyperfine structure lines were summed over all components using their statistical weight (see Bruderer et al. 2010b for details). The observed parameters of H$_3$O$^+$ at 1031 GHz and 1069 GHz  are extracted from blends (see Fig. 1) by fitting Gaussians.

A rotational diagram  can be constructed for H$_3$O$^+$ (see Appendix). The data can be fitted with a straight line, suggesting that {\it (i)} the ortho-to-para ratio is compatible with unity and {\it (ii)} a unique rotational temperature of 239 K (error range 205 - 285 K) exists. It indicates that the H$_3$O$^+$ emission originates, on weighted average, from regions with equal or even hotter kinetic temperature, or, if pumped, radiation temperature.

\begin{figure}[htb]
\centering
\resizebox{6cm}{!}{\includegraphics{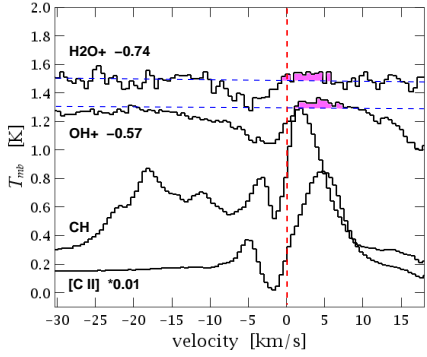}}
\caption{Light hydrides with P Cygni type absorption features compared with [C II]. The flux scales are reduced and shifted for better visibility. The velocity is relative to the systemic velocity of W3 IRS5, shown by a vertical dashed line. The horizontal dashed lines indicate the continuum level.}
\label{PCyg}
\end{figure}

\section{Discussion and conclusions}
The observed shifts of the line peaks relative to the systemic velocity of the YSO are small and suggest that the lines originate in the star-forming region, not in the foreground interstellar medium. This may not be the case, however, for the components OH$^+$ and H$_2$O$^+$ shifted by a larger amount to roughly to the velocity $V_{LSR}=0$.

Additional evidence of the origin comes from the lines in emission, which are produced at critical densities (if known) of order $> 10^{7}$cm$^{-3}$. Even lines predominantly in absorption, such as OH$^+$ and possibly H$_2$O$^+$, have a small emission feature in the red wing, thus a P-Cyg-like profile, near the systemic velocity. This is the first report of Galactic OH$^+$ and H$_2$O$^+$ in emission. Figure 2 shows the similarity of the profiles of OH$^+$ and H$_2$O$^+$. The absorption in CH may be caused by self-absorption, but is similar to that in [C II], which is found in emission at the off position. Absorption lines in star-forming regions need detailed modeling and radiation transfer calculations beyond the scope of this letter.

\begin{figure}[htb]
\centering
\resizebox{7cm}{!}{\includegraphics{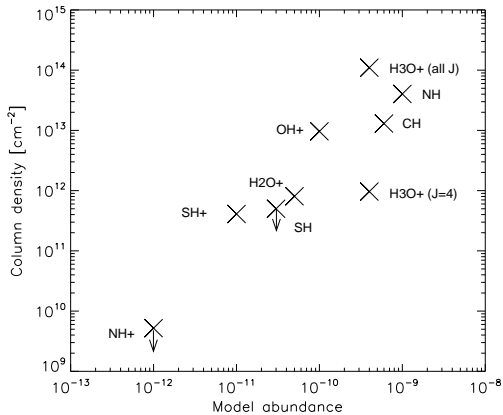}}
\caption{Comparison of low-$J$ level column density with predicted beam-averaged abundance relative to H$_2$ from a 2D model by Bruderer et al. (2010a). }
\label{Comparison}
\end{figure}

The ground-state column densities of the component moving with the YSO are $8.1\times 10^{11}$ cm$^{-2}$ for H$_2$O$^+$, and $9.7\times 10^{12}$ cm$^{-2}$ for OH$^+$. The values for the red-shifted components are $3.8\times 10^{12}$ cm$^{-2}$ and $6.1\times 10^{13}$ cm$^{-2}$, respectively. Both components yield larger OH$^+$/H$_2$O$^+$ ratios than the other observations reported in this volume.

The measured line widths are generally small ($<$ 7 km s$^{-1}$) and show no anisotropies. We thus find no evidence of shocks except possibly in OH$^+$, which needs to be studied in combination with shock tracers.

In Fig. 3, column densities are displayed, derived from integrated line fluxes neglecting re-emission or reabsorption of the final state. The derivation is based on Table 1, except for OH$^+$ and H$_2$O$^+$ where only the unshifted component attributed to the YSO is used. For H$_3$O$^+$, the value of the $J=4$ level as well as the one from the rotational diagram, integrated over all levels, are shown. The column densities are compared with abundances predicted by Bruderer et al. (2010a) in a two-dimensional `standard' YSO model used here as a template, assuming UV and X-ray irradiation by a central high-mass YSO. It enhances the abundance of diatomic hydrides in the outflow walls by many orders of magnitude, such that the beam-averaged abundance is significantly changed. It is averaged over a radius of 20 000 AU, or 10.9$''$ at the distance of W3 IRS5.

The similarity of observations and model abundances in Fig. 3 support the scenario of hydride enrichment in outflow walls heated and irradiated by protostellar far UV.

\begin{acknowledgements}
We thank Michael Kaufman and Serena Viti for helpful comments on an early draft. This program is made possible thanks to the Swiss {\it Herschel} guaranteed time program. HIFI has been designed and built by a consortium
of institutes and university departments from across Europe, Canada and the United States under the leadership of SRON Netherlands Institute for Space Research Groningen, The Netherlands and with major contributions from Germany, France, and the US. Consortium members are: Canada:CSA, U.Waterloo; France: CESR, LAB, LERMA, IRAM; Germany: KOSMA,MPIfR, MPS; Ireland: NUI Maynooth; Italy: ASI, IFSI-INAF, Osservatorio Astrofisico di Arcetri-INAF; Netherlands: SRON, TUD; Poland: CAMK, CBK; Spain: Observatorio Astronomico Nacional (IGN), Centro de Astrobiologia (CSIC-INT); Sweden: Chalmers University of Technology, Onsala Space Observatory, Swedish National Space Board, Stockholm University; Switzerland: ETH Zurich, FHNW; USA: Caltech, JPL, NHSC. The work on star formation at ETH Zurich is partially funded by the Swiss National Science Foundation (grant nr. 200020-113556).
\end{acknowledgements}

\Online

\begin{appendix}

\section{}

The rotational diagram in Fig. A1 is complemented with ground-based data for the frequency range 300 - 400 GHz observed at the CSO with comparable beam size by Phillips, van Dishoeck \& Keene (1992). The non-detection at 307 MHz is surprising, but possibly an effect of optical depth. The data in Fig. A.1, except 307 GHz, are well fitted by a single rotational temperature of 239 K, suggesting that the observed levels are populated according to an exponential distribution. The derived temperature and column density infer an optical depth of $\tau < 0.1$ for all lines except at 307 GHz. The fitted line (dashed) corresponds to a column density of 8.5$(\pm 2)\times 10^{13}$ cm$^{-2}$, consistent with the value derived by Phillips, van Dishoeck \& Keene (1992). This leads to a beam averaged H$_3$O$^+$ abundance of 4.2$(\pm 1)\times 10^{-10}$ relative to H, to be compared with the theoretical value of 4 $\times 10^{-10}$ reported by Bruderer et al. (2010a).

\begin{figure}
\centering
\resizebox{\hsize}{!}{\includegraphics{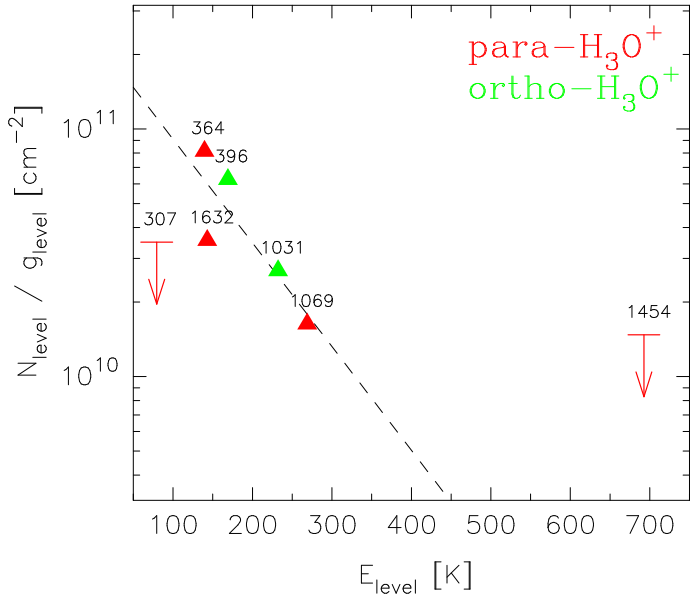}}
\caption{Rotational diagram of H$_3$O$^+$. Numbers indicate the frequencies in GHz of the observed lines. }
\label{Comparison}
\end{figure}

\end{appendix}

\end{document}